# Rapid green biosynthesis of Silver–montmorillonite nanocomposite using water extract of Satureja hortensis L and evaluation of the antibacterial properties

**Sajjad Sedaghat[a] and Mehdi Khalaj[b]**

[a]Department of Chemistry, Faculty of Science, Shahr-e-Qods Branch, Islamic Azad University, Tehran, Iran
[b]Young Researchers and Elite Club, Buinzahra Branch, Islamic Azad University, Buinzahra, Iran

**Keywords:** Antibacterial effects; Montmorillonite; Nanocomposite; Satureja hortensis l; Transmission electron microscopy

In this work, facile, rapid and green biosynthesis and characterization of silver–montmorillonite (MMT) nanocomposite is reported at room temperature. Silver nanoparticles (Ag–NPs) were prepared by water extract of Satureja hortensis (L) as the reducing agent and MMT as interlamellar space for controlling the size of Ag-NPs. MMT was suspended in the aqueous $AgNO_3$ solution, and after the absorption of silver ions, $Ag^+$ was reduced to $Ag°$ using water extract of Satureja hortensis (L). Evaluations of the antibacterial properties are also reported. The nanocomposite was characterized by ultraviolet–visible spectroscopy (UV–Vis), powder X–ray diffraction (XRD), scanning electron microscopy (SEM) and transmission electron microscopy (TEM). TEM study showed the formations of nanocomposite using water extract of Satureja hortensis L in the 4.88 – 26.70 nm range and an average particles size of 15.79 nm. In addition, XRD study showed that the particles have a face centered cubic (FCC) structure. The nanocomposite showed antibacterial properties against Gram positive and negative bacteria.

## 1. Introduction

Nanoparticles (NPs) have recently attracted much attention. Therefore, the synthesis of nano silver (Ag-NPs) has been a subject of great interest in the past several years. The obtained Ag-NPs have unique electrical, optical and biological properties. Ag-NPs normally have short lives in solution and may agglomerate fast. .To prevent this nanoparticles can be supported between the internal spaces of clay or on its external surfaces [1,2]. Thus, the synthesis of Ag-NPs onto MMT supports with swelling and ion exchange properties is a good way to control the particle size. In particular, the interlayer space of MMT can be used for the synthesis of metal NPs as the support for cations, as discussed in our previous Works [2-5,12]. The synthesis of $TiO_2$-montmorillonite nanocomposites into the interlayers of montmorillonite has also been successfully carried out[6]. In another study, silver nanoparticles (Ag-NPs) were prepared using chemical reduction into the interlayer of montmorillonite[7]. Silver-montmorillonite-chitosan nanocomposites were also synthesized using UV irradiation as the reducing agent, chitosan as the stabilizer and montmorillonite as a solid support[8]. Synthesis and characterization of silver nanoparticles (Ag-NPs) using UV irradiation to $AgNO_3$ in the interlayer of montmorillonite without any reducing agent or heat treatment was also reported in another work[9,10]. Biomolecules presents in plant water extracts can be used to reduce silver ions to Ag-NPs as a green chemical, via a cost effective and environmentally friendly route[11]. In our previous work, silver nanoparticles were synthesized with an average size of 15 ± 7.402 nm and spherical shape using the water extract of Satureja hortensis L at room temperature without any support[12]. In this article, a simple and rapid green bioreduction method for synthesizing Ag-NPs in the interlayer space of MMT is reported. This method consists of controlled reduction without any undesired oxidation products. Ag-NPs were intercalated into the lamellar space of MMT using water extract of Satureja hortensis L in the absence of heat treatment or chemical reducing agents. MMT was used as the protective colloid to prevent Ag-NPs from aggregation. The antibacterial effect of the nanocomposite was evaluated against Escherichia coli (Gram negative) and Staphylococcus aureus (Gram positive) bacteria by the minimum bactericidal concentration (MBC) method.

*Author for correspondence sajjadsedaghat@yahoo.com).
†Present address: Department of Chemistry, Faculty of Science, Shahr-e-Qods Branch, Islamic Azad University, Tehran, Iran

## 2. Materials and Methods
### 2.1. Instruments and reagents

High-purity chemical reagents were purchased from the Merck and Aldrich chemical companies. All materials were of commercial reagent grade and were used without further purification. Doubly distilled water was used in all experiments. FT-IR spectra were recorded on a Nicolet 370 FT/IR spectrometer (Thermo Nicolet, USA) using pressed KBr pellets. X-ray diffraction (XRD) measurements were carried out using a Philips powder diffractometer type PW 1373 goniometer (Cu K$\alpha$ = 1.5406 A°). The scanning rate was 2°/min in the 2θ range from 10 to 80°. Scanning electron microscopy (SEM) was performed on a Cam scan MV2300. The shape and size of the TMAN were identified by transmission electron microscope (TEM) using a Philips EM208 microscope operating at an accelerating voltage of 90 kV. The UV–visible spectra were also recorded using UV Bio–TEK UV–visible spectrophotometer from 325 nm to 800 nm. The antibacterial activity of the synthesized nanocomposite, kanamycin and vancomycin were evaluated against Gram negative bacteria (Escherichia coli ATCC 25922) and Gram positive bacteria (Staphylococcus aureus ATCC 25923) using the minimum bactericidal concentration (MBC) method. The interlamellar spacing variations for MMT and Ag/MMT nanocomposites were studied in the angle range of 2°, 2θ, and 12°. The interlamellar space was determined based on the XRD peak positions using Bragg's equation. A wave–length ($\lambda$) of 0.15418 nm was used for this measurements[13-15].

### 2.2. Preparing the water extract of Satureja hortensis L

Fresh leaves of Satureja hortensis L were collected from Shahryar region of Iran in November 2014. $AgNO_3$ (99.80%), used as the silver source, was purchased from Merck Chemical Co. and the montmorillonite powder (MMT), used as the solid support, was purchased from Fluka Chemical Co. Kanamycin (antibiotic) was supplied by Duchefa Biocheme and Vancomycin (antibiotic) was obtained from Sigma Aldrich Chemical Co. Double distilled water was used in all the experiments. Fresh leaves of Satureja hortensis L were washed and dried for a week in shade. The Satureja hortensis L dried leaves were extracted in water and used for reduction of $Ag^+$ ions to $Ag°$. For this purpose, 10 g of dried leaves of the plant were added to 100 mL of double distilled water and the mixture was boiled for 10 minutes in a water bath. The extract was then filtered and centrifuged at 4000 rpm for 10 minutes to remove all proteins and other residues from the extract.

### 2.3. Synthesis of silver nanoparticles

1.00 g of MMT powder was dispersed with vigorous stirring in certain amount of double distilled water for 1 hour. MMT suspension was added to 100 mL of 0.01 (mol $L^{-1}$) $AgNO_3$ solution for the synthesis of Ag/MMT nano composite. The mixture was then added to 20 mL of Satureja hortensis L water extract at room temperature while sonicating and then vigorously stirred for 48 hours.

## 3. Results and Discussion
### 3.1. UV absorption measurements

In Figure 1, the UV-visible absorption spectra of MMT based on Ag-NPs prepared by water extract of Satureja hortensis L, MMT and plant extract are shown (a-c). The characteristic Ag SPR bands were detected in the range of 360-415 nm. These absorption bands were due to the Ag-NPs smaller than 30 nm.16 The absorption peak in the wavelength of 400 nm confirmed the formation of Ag/MMT nanocomposite in nanometer size in correlation to TEM results as 15.79 nm.

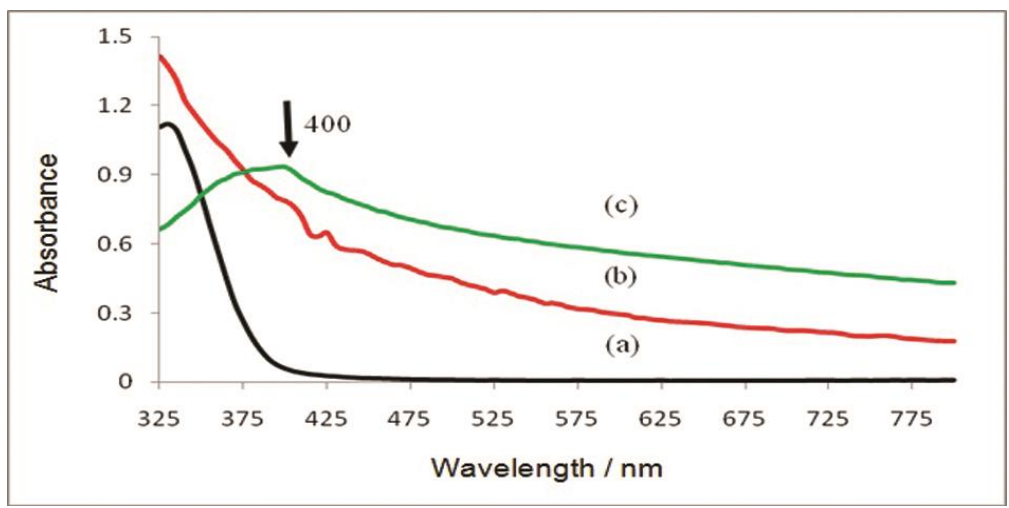

**Figure 1**. UV-Vis absorption spectra: (a) plant extract of Satureja hortensis L, (b) (MMT) and (c) Ag/MMT nanocomposite suspension in water after 48 hours.

## 3.2. X-ray diffraction (XRD) patterns

The X-ray diffraction (XRD) patterns of the pure montmorillonite and the synthesized Ag/MMT nanocomposite are shown in Figure 2. In the Ag/MMT nanocomposite, the XRD peaks at 2θ of 38.19°, 44.38°, 64.44° and 77.51° correspond to (111), (200), (220) and (311) planes of the face centered cubic (FCC) of nanocomposite, respectively. The mean particles sizes of nanocomposites were evaluated using the Debye-Scherer Equation 1:

$$d = \frac{K\lambda}{\beta cos\theta} \qquad (1)$$

Where λ is the X-ray wavelength (1.540560 Å), β is the width of the XRD peak, θ is the Bragg angle, d is the particles size and K is the Scherer constant with a value 0.9. The average sizes of nanocomposites were calculated to be about 16 nm.

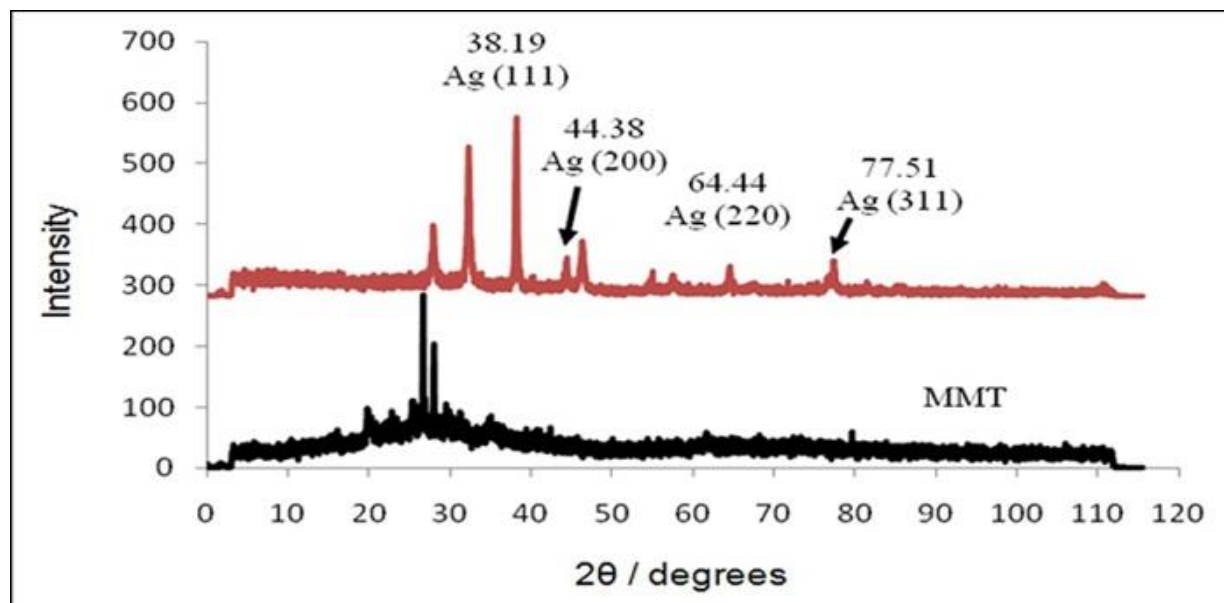

**Figure 2**. XRD patterns of montmorillonite (MMT) and silver-montmorillonite nanocomposite synthesized by Satureja hortensis L water extract after 48 hours.

## 3.3. SEM and TEM observations

SEM image of Ag/MMT nanocomposite, prepared by green reduction of AgNO3 in montmorillonite using water extract of Satureja hortensis L at room temperature, is shown in Figure 3. SEM images were used to study the morphology of Ag/MMT nanocomposite. The shape of the particles was found to be spherical and their average size was about 16 nm.

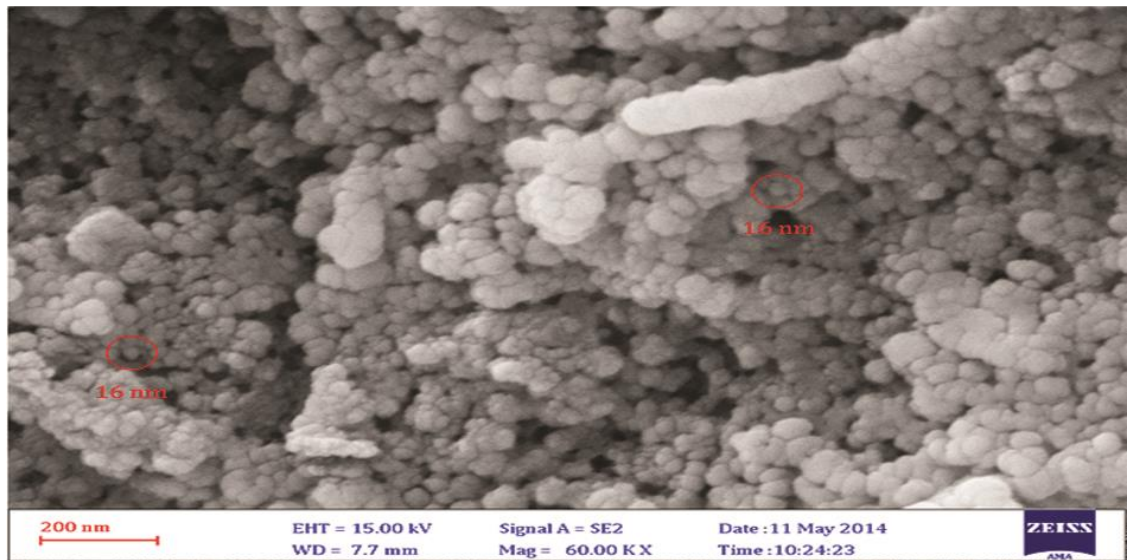

**Figure 3.** SEM image of Ag/MMT nanocomposite after 48 hours.

The formation of Ag/MMT nanocomposite suspension was confirmed by the TEM image. Figures 4a and 4b show the TEM images of montmorillonite and nanocomposite, resepctively, and Figure 4c shows the size distribution of the Ag/MMT nanocomposite suspension. TEM image and the size distribution of Ag/MMT nanocomposite suspension show the mean diameter and $15.79 \pm 10.91$ nm standard deviation of the nanocomposite for particles prepared in the montmorillonite layers. In addition, the TEM image confirmed the presence of a layer surrounding the nanocomposite, which prevented the agglomeration of the particles. These layer surroundings are depicted in the figure with black arrows.

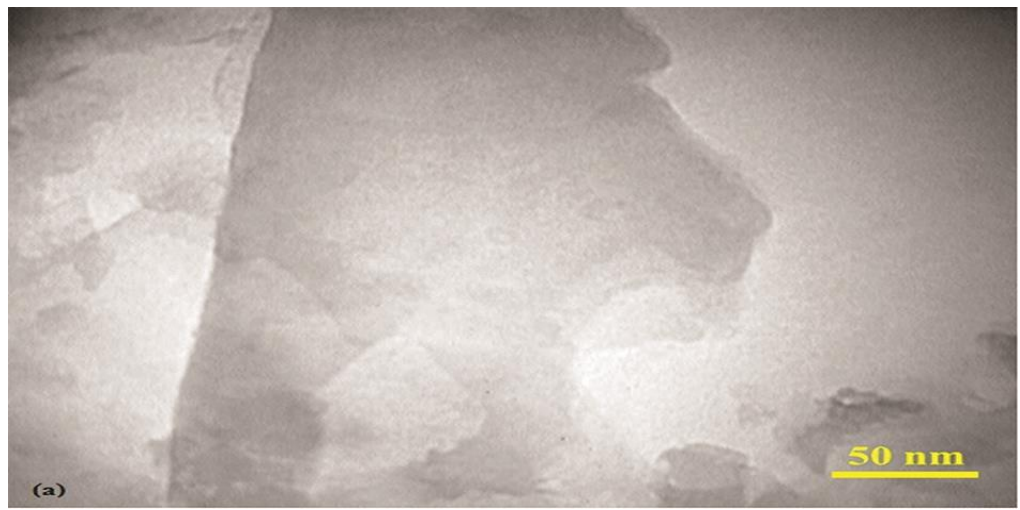

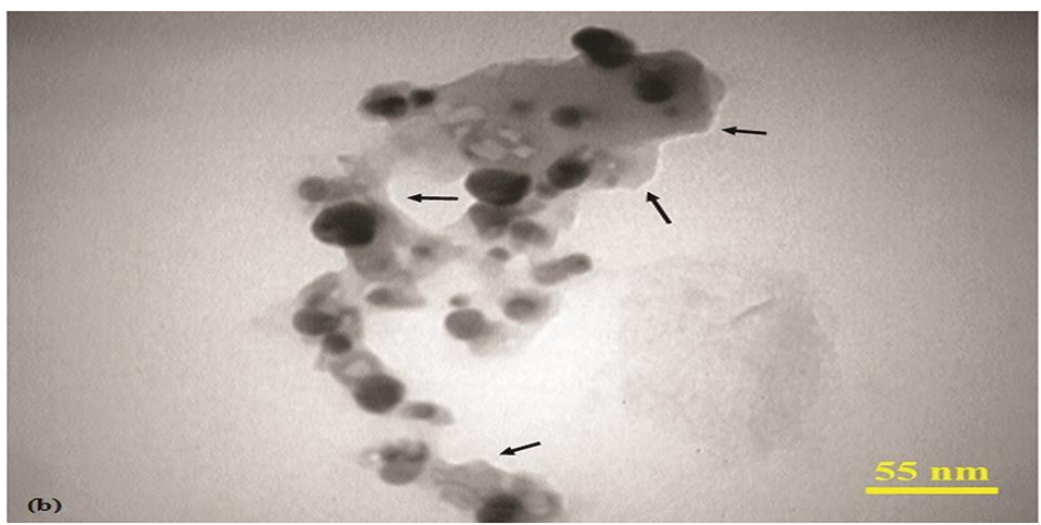

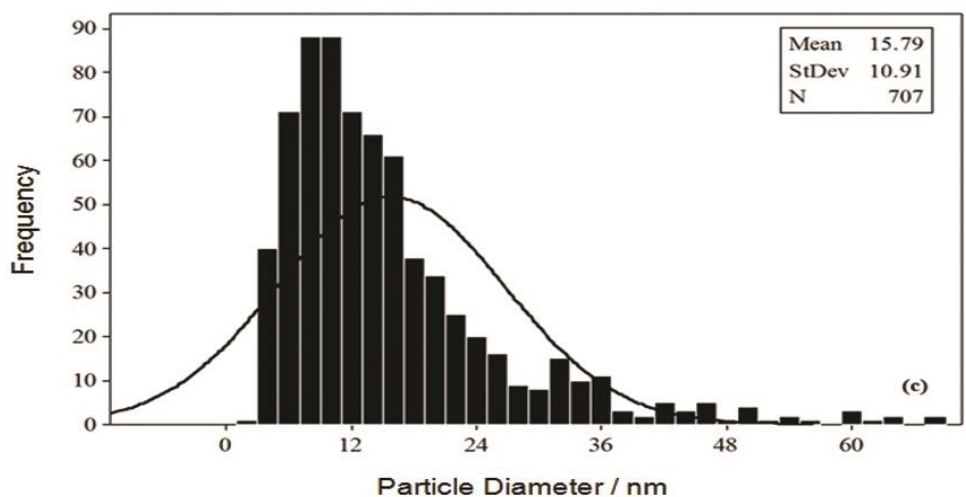

**Figure 4.** TEM image: (a) montmorillonite, (b) Ag/MMT nanocomposite and (c) corresponding size of Ag/MMT nanocomposite after 48 hours.

## 3.4. FT–IR spectra

Figure 5 Shows the FT-IR spectra of MMT, plant extract and Ag/MMT nanocomposite. In MMT, the vibration band at 3634 cm-1 corresponds to O–H stretching and the peak at 3442 cm-1 is due to the interlayered O–H stretching (H-bonding). The vibration band at 634 cm-1 is associated with Al–OH. The interactions between MMT and Ag-NPs are indicated by the peak at 3442 cm-1. The broad peak is related to the presence of van der waals interactions between the hydroxyl groups of MMT layers and the partial positive charge of the surface of Ag-NPs.9 FT-IR spectroscopy was carried out to determine the potential biomolecules responsible for the reduction and capping of the silver nanoparticles synthesized. The FT-IR spectra of plant extract shows absorption bands characteristics of functional groups such as alcohol, phenol, amine and carbonyl group.

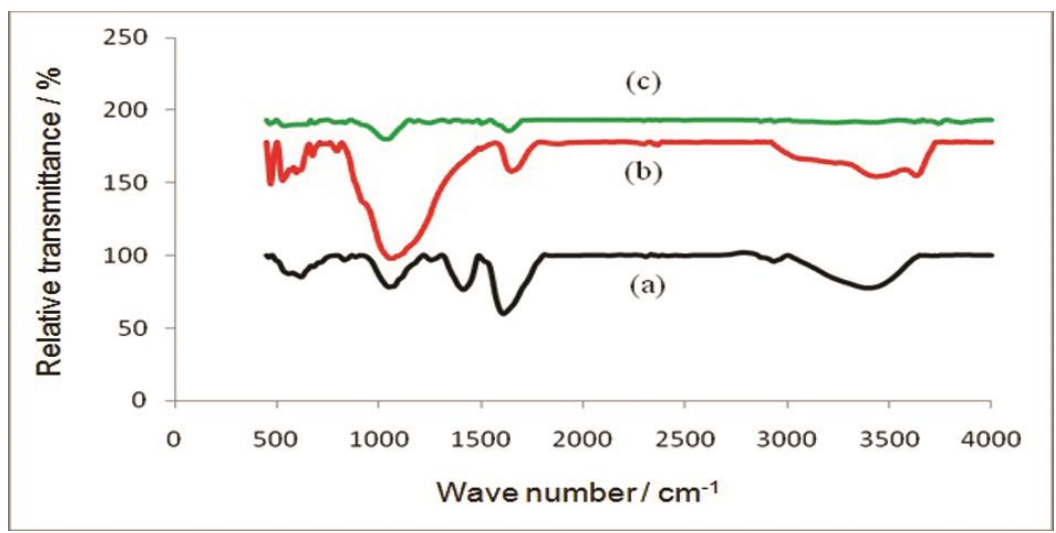

**Figure 5.** FT–IR spectra: (a) plant extracts of Satureja hortensis L, (b) montmorillonite and (c) Ag/MMT nanocomposite after 48 hours.

## 3.5. Antibacterial tests

The antibacterial tests of Ag/MMT nanocomposite and selected antibiotics against Escherichia coli and Staphylococcus aureus bacteria were successfully carried out. The result showed that the effective concentration of Kanamycin antibiotic (MBC) against Escherichia coli, effective concentration of Vancomycin antibiotic (MBC) against Staphylococcus aureus and effective concentration of Ag/MMT nanocomposite (MBC) against Staphylococcus aureus and Escherichia coli bacteria were 31.25 ($\mu$g mL$^{-1}$), 4 ($\mu$g mL$^{-1}$), 0.193 ($\mu$g mL$^{-1}$) and 0.193 ($\mu$g mL$^{-1}$), respectively.

Table 1. The values of initial and effective concentrations of antibiotics and silver-montmorillonite nanocomposite against Escherichia coli and Staphylococcus aureus bacteria

| Compound | Variable | | |
|---|---|---|---|
| | The minimum bactericidal concentration (MBC) dilution | Initial concentration (μg mL$^{-1}$) | Effective concentration (μg mL$^{-1}$) |
| Kanamycin antibiotic against *Escherichia coli* bacteria | 1:32 | 1000 | 31.25 |
| Vancomycin antibiotic against *Staphylococcus aureus* bacteria | 1:64 | 256 | 4 |
| Silver-montmorillonite nanocomposite against *Escherichia coli* bacteria | 1:4096 | 792 | 0.193 |
| Silver-montmorillonite nanocomposite against *Staphylococcus aureus* bacteria | 1:4096 | 792 | 0.193 |

# 4. Conclusions

In summary, Ag/MMT nanocomposites were successfully synthesized by a facile and green method using water extract of Satureja hortensis L at room temperature and their antibacterial properties were evaluated. The size of nanocomposite particles was between 4.88 and 26.70 nm. The Ag/MMT nanocomposites were prepared via an eco–friendly biological approach without using any toxic chemicals as the reducing agent. The formation of Ag/MMT nanocomposite was confirmed by UV-Visible absorption spectra, TEM, SEM images and the X–ray diffraction (XRD) patterns. The TEM image showed that the Ag/MMT nanocomposites were of spherical shape and X–ray result confirmed that the obtained Ag/MMT nanocomposites have a face centered cubic crystal structure. The antibacterial tests of Ag/MMT nanocomposites on Escherichia coli (Gram negative) and Staphylococcus aureus (Gram positive) bacteria by the minimum bactericidal concentration (MBC) method were successfully carried out and the results showed that Ag/MMT nanocomposite had strong antibacterial properties.

**Tables**

**Table 1.** The values of initial and effective concentrations of antibiotics and silver-montmorillonite nanocomposite against Escherichia coli and Staphylococcus aureus bacteria

| Compound | Variable | | |
| --- | --- | --- | --- |
| | The minimum bactericidal concentration (MBC) dilution | Initial concentration ($\mu g\ mL^{-1}$) | Effective concentration ($\mu g\ mL^{-1}$) |
| Kanamycin antibiotic against *Escherichia coli* bacteria | 1:32 | 1000 | 31.25 |
| Vancomycin antibiotic against *Staphylococcus aureus* bacteria | 1:64 | 256 | 4 |
| Silver-montmorillonite nanocomposite against *Escherichia coli* bacteria | 1: 4096 | 792 | 0.193 |
| Silver-montmorillonite nanocomposite against *Staphylococcus aureus* bacteria | 1:4096 | 792 | 0.193 |

**Figures**